# Mass, Luminosity, and Stellar Age of Early-type Stars from the LAMOST Survey

Qida Li[1,2], Jianping Xiong[1], Jiao Li[1], Yanjun Guo[1], Zhanwen Han[1], Xuefei Chen[1,3], and Chao Liu[2,4]
[1] Yunnan Observatories, Chinese Academy of Sciences, Kunming, 650216, People's Republic of China; cxf@ynao.ac.cn
[2] School of Astronomy and Space Science, University of Chinese Academy of Sciences, Beijing, 100049, People's Republic of China
[3] International Centre of Supernovae, Yunnan Key Laboratory, Kunming, 650216, People's Republic of China
[4] Key Laboratory of Space Astronomy and Technology, National Astronomical Observatories, Chinese Academy of Sciences, Beijing 100101, People's Republic of China

## Abstract

Mass ($M$) and luminosity ($L$) are fundamental parameters of stars but can only be measured indirectly. Typically, effective temperature ($T_{\rm eff}$), surface gravity ($\log g$), and metallicity ([M/H]) are derived from stellar spectra, then $M$, $L$, and stellar age ($t$) can be obtained by interpolating in the grid of stellar evolutionary models. In this paper, we use random forest in combination with the evolutionary grid from PARSEC 1.2S to determine $M$, $L$, $t$, and initial mass ($M_{\rm i}$) for early-type main-sequence stars ($T_{\rm eff} \geqslant 7000$ K) as identified from the LAMOST survey. The convex hull algorithm is employed to select the main-sequence stars. The results demonstrate that the prediction precision is 32% and 9% for $L$ and $M$, respectively, which is comparable to that achieved by fitting evolutionary tracks. Furthermore, the predicted $L$ also aligns with Gaia's observations, with a relative difference of 36%. The prediction for $t$ is relatively less accurate, indicating a difference of 0.44 Gyr for two components in wide binaries. This discrepancy is due to the inconsistent metallicity measurements. For the two sets of atmospheric parameters we used, the relative differences in $L$, $M$, $t$, and $M_{\rm i}$ are 29%, 7%, 36%, and 7%, respectively. The influence of metallicity on these parameters is analyzed, with the conclusion that metallicity has the greatest impact on $t$. Consequently, two catalogs are presented, which could be useful for studying stellar populations such as the luminosity function and initial mass function of early-type stars.

*Unified Astronomy Thesaurus concepts:* Catalogs (205); Early-type stars (430); Random Forests (1935)

*Materials only available in the online version of record:* tar.gz file

## 1. Introduction

Early-type stars are generally known as O-, B-, and A-type stars in the spectral hierarchy. They are massive, luminous, and evolve rapidly (D. C. Morton & T. F. Adams 1968; N. Panagia 1973). These stars are excellent candidates for studying the structures and dynamic processes of star-forming regions and the Galaxy (G. Carraro et al. 2010, 2017; Y. Xu et al. 2018; Gaia Collaboration et al. 2023). Massive early-type stars are also progenitors of various celestial objects, including neutron stars, black holes (A. Sadowski et al. 2008; N. Langer et al. 2020), gamma-ray bursts (S.-C. Yoon & N. Langer 2005; S. E. Woosley & A. Heger 2006), and supernova (N. Langer et al. 2007; C. Inserra & S. J. Smartt 2014). Moreover, stripped massive stars, either resulting from strong stellar-wind loss (Wolf–Rayet stars, N. Moens et al. 2022; J. N. Rustamov & A. F. Abdulkarimova 2023) or from binary evolution (massive helium stars), played a significant role in ionizing neutral hydrogen during the early stages of the Universe (Y. Götberg et al. 2020). The fundamental characteristics of early-type stars, such as mass ($M$), radius ($R$), luminosity ($L$), and age ($t$), as well as effective temperature ($T_{\rm eff}$), surface gravity ($\log g$), and metallicity ([M/H]), are essential for our understanding of their formation, evolution, and impact on celestial objects. These parameters are also crucial for the study of stellar populations and their role in the early Universe.

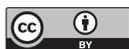

In general, the atmospheric parameters of stars (i.e., $T_{\rm eff}$, $\log g$, [M/H]) can be measured directly through stellar spectra (Y.-S. Ting et al. 2017a, 2017b; M. Xiang et al. 2019). However, accurate measurements of stellar mass and age are required for stellar models or photometric observations. For example, high-precision estimations of stellar masses can be obtained through the orbital solutions of binaries (D. R. Soderblom 2010; M. Xiang et al. 2017; M. Dai et al. 2022; T. Chen et al. 2023; J. Xiong et al. 2023) or asteroseismology (N. Gai et al. 2011; W. J. Chaplin et al. 2014; M. H. Pinsonneault et al. 2018; A. Serenelli et al. 2021). The application of these methods to high-mass stars is quite limited because only a few such stars have been observed with long-term and high-precision data (light curves and spectra). The fitting of stellar evolutionary tracks or isochrones has been widely used for determining stellar masses and ages, especially for massive stars (A. Serenelli et al. 2021). The strong nonlinearity of stellar evolutionary tracks or isochrones makes the process of obtaining parameters a bit cumbersome and inefficient. Moreover, the degeneracy of stellar age and metallicity make it difficult to accurately determine the two parameters. B. R. Jørgensen & L. Lindegren (2005) introduced a Bayesian approach in the isochrone fitting, which has been used in several studies (F. Pont & L. Eyer 2004; B. R. Jørgensen & L. Lindegren 2005; Z. Shkedy et al. 2007).

With the boosting of large survey projects, machine learning techniques have been used in astronomic data mining and parameter determination due to their accuracy and efficiency. They have become an indispensable tool in classifying objects (Y.-S. Ting et al. 2018; Y. Guo et al. 2024), deriving atmospheric parameters and elemental abundances from spectra (R. Wang et al. 2020; B. Zhang et al. 2020; Y. Guo et al. 2021;

W. Sun et al. 2021; M. Xiang et al. 2022), as well as predicting luminosity, mass, and age for turn-off stars, giant stars, and clusters (Y. Wu et al. 2018; T. J. Armitage et al. 2019; V. Bonjean et al. 2019; P. Das & J. L. Sanders 2019; Y. Bu et al. 2020; M. Huertas-Company et al. 2020; N. Gupta & C. L. Reichardt 2021; D. Kodi Ramanah et al. 2021; S. Mucesh et al. 2021). For red clump stars, the predicted precision of mass and age via machine learning can reach around 9% and 18%, respectively (Q.-D. Li et al. 2022).

The Large Sky Area Multi-Object Fiber Spectroscopic Telescope (LAMOST) has been in operation for over a decade and has released more than 20 million stellar spectra (X.-Q. Cui et al. 2012; L.-C. Deng et al. 2012; X.-W. Liu et al. 2014). A large sample of early-type stars have been observed by the LAMOST survey and their atmospheric parameters ($T_{\text{eff}}$, $\log g$, [M/H]) have been measured from the spectra. For example, W. Sun et al. (2021) derived $T_{\text{eff}}$, $\log g$, and [M/H] for late-B and A-type stars from LAMOST medium-resolution spectra with a precision of 75 K, 0.06 dex, and 0.05 dex, respectively. Y. Guo et al. (2021) obtained the atmospheric parameters for early-type (O/B/A) stars from LAMOST low-resolution spectra with a precision of 1642 K and 0.25 dex for $T_{\text{eff}}$, $\log g$, respectively, comparing the results based on high-resolution spectra. M. Xiang et al. (2022) also provided a catalog of atmospheric parameters for 332,172 stars with $T_{\text{eff}} \geqslant 7000$ K. These parameters were determined using the spectra from LAMOST DR6. And for a typical B-type star with signal-to-noise ratio (S/N) = 100, the precision for the $T_{\text{eff}}$, $\log g$, and [Fe/H] is 300 K, 0.03 dex, and 0.1 dex, respectively. It is possible to create a uniform catalog for early-type stars that contains the stellar mass, luminosity, and age based on the atmospheric parameters. This catalog will be useful for studying the evolution of early-type stars and related celestial objects, as well as for exploring massive stellar populations and their role in cosmology.

In this paper, we utilize machine learning techniques to measure the $L$, $M$, $t$, and initial mass ($M_i$) of early-type stars in the LAMOST survey. As there are only a few observed early-type stars with measurements of both stellar mass and age (P. Massey et al. 2005; M. M. Hohle et al. 2010), we use theoretical evolutionary tracks as mock data to build our model, then apply our model to the observed data. The paper is structured as follows. We first present the atmospheric parameters of catalogs and the mock data we used in Section 2. We then give a brief introduction to the machine learning method, as well as the training process and the convex hull algorithm in Section 3. In Section 4, we present the prediction results. The validation and the impact of metallicity have been discussed in Section 5. Finally, we give a summary in Section 6.

## 2. Data

### 2.1. The Atmospheric Parameters

Two catalogs containing early-type stars with atmospheric parameters obtained from LAMOST are utilized in this work. The samples comprise a total of 16,032 stars (22,292 spectra) from Y. Guo et al. (2021) and 332,172 stars (454,693 spectra) from M. Xiang et al. (2022). To ensure the accurate measurement of atmospheric parameters, we performed quality control on the data from the two star catalogs, as detailed below.

**Table 1**
The Credible Range of the Two Catalogs

| Parameter | Y. Guo et al. (2021)[a] | M. Xiang et al. (2022)[b] |
|---|---|---|
| Total number of stars | 16,032 | 332,172 |
| $T_{\text{eff}}$ (K) | 15,000 ∼ 55,000 | 7500 ∼ 60,000 |
| $\log g$ (cm s$^{-2}$) | 1.75 ∼ 4.75 | 0 ∼ 5 |
| Metallicity (dex) | −1 ∼ 0.3 | −3 ∼ 0.5 |

**Notes.**
[a] Obtained from the LAMOST low-resolution survey by Stellar LAbel Machine (SLAM; B. Zhang et al. 2020).
[b] Derived from LAMOST low-resolution spectra through the HotPayne method.

First, we performed an initial filtering of the samples based on the credible ranges; the credible ranges of atmospheric parameters for the two catalogs are presented in Table 1. We then select spectra with an S/N greater than 15, and for stars with multiple observations, we retain only the one with the highest S/N. Our study focuses on early-type stars, which are typically massive and young. These stars are expected to form from the local interstellar medium, which is approximately solar in metallicity. This is supported by studies of the local age–metallicity relation (L. Casagrande et al. 2011; I. Minchev et al. 2018), showing that stars just a few Gyr old typically have [M/H] values between −0.5 and 0.5 dex. Figure 1 shows the [M/H] of samples with S/N > 15 from the catalogs of M. Xiang et al. (2022) and Y. Guo et al. (2021). It is evident that only a small number of samples have [M/H] values exceeding the range of [−0.5, 0.5]. And by reviewing the samples, we find that the metal-poor stars have larger measurement uncertainties. Therefore, we have restricted the [M/H] range of the observed samples to [−0.5, 0.5]. Finally, we obtain 4254 stars from Y. Guo et al. (2021) and 166,510 stars from M. Xiang et al. (2022) for making predictions.

### 2.2. Mock Data

As there is a limited sample of early-type stars with accurate measurements for $L$, $M$, and $t$ in observations, it is insufficient for the purpose of training machine learning models. We thus use the stellar evolutionary model PARSEC 1.2S (A. Bressan et al. 2012; Y. Chen et al. 2014, 2015; X. Fu et al. 2018) to construct our mock data. The $T_{\text{eff}}$, $\log g$, and [M/H] are typically derived from the observational spectra. The $M$ ($M = gL/4\pi\sigma GT_{\text{eff}}^4$, $\sigma$ is the Stefan–Boltzmann constant) can be calculated by comparing the observed position on a $T_{\text{eff}}$–$\log g$–$\log L$ diagram with a grid of evolutionary tracks generated for different metallicities (A. Serenelli et al. 2021). Similarly, $L$ can also be determined by comparing their position on a $T_{\text{eff}}$–$\log g$–[M/H] diagram with a grid of evolutionary tracks. Once $L$ and $M$ are known, we can subsequently determine the $t$. Consequently, we extract the theoretical $T_{\text{eff}}$, $\log g$, [M/H], $\log L$, $M$, and $t$ from the PARSEC 1.2S evolutionary tracks to construct the mock data. Figure 2 displays the evolutionary tracks of PARSEC 1.2S on the $T_{\text{eff}}$–$\log g$ relation. The left panel shows the evolutionary tracks with different $M_i$ for a metallicity of $Z = 0.02$ and helium abundance of $Y = 0.284$. For this work, only the main-sequence tracks (i.e., the central hydrogen mass fraction $X_c$ in a range of 0.01% ∼ 99% of the maximum value, the orange solid lines in the left panel) from the PARSEC 1.2S are selected as the mock data because we focus on the main-sequence stars. The right panel shows selected evolutionary tracks from PARSEC 1.2S on the $T_{\text{eff}}$–$\log g$ diagram.

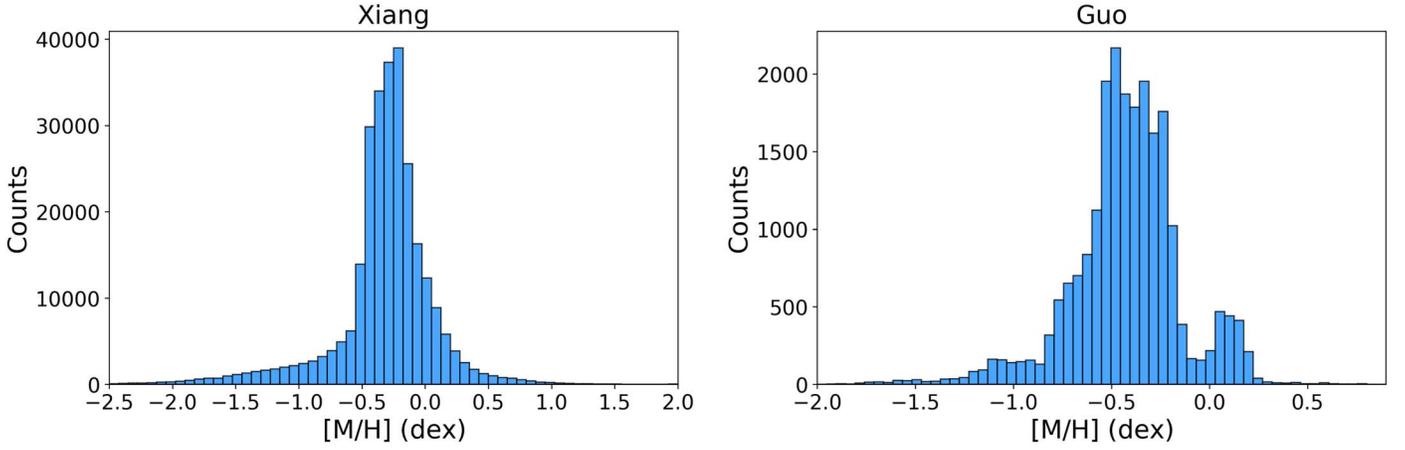

**Figure 1.** The distribution of [M/H] in the catalogs of M. Xiang et al. (2022) (left) and Y. Guo et al. (2021) (right), showing only stars with S/N greater than 15.

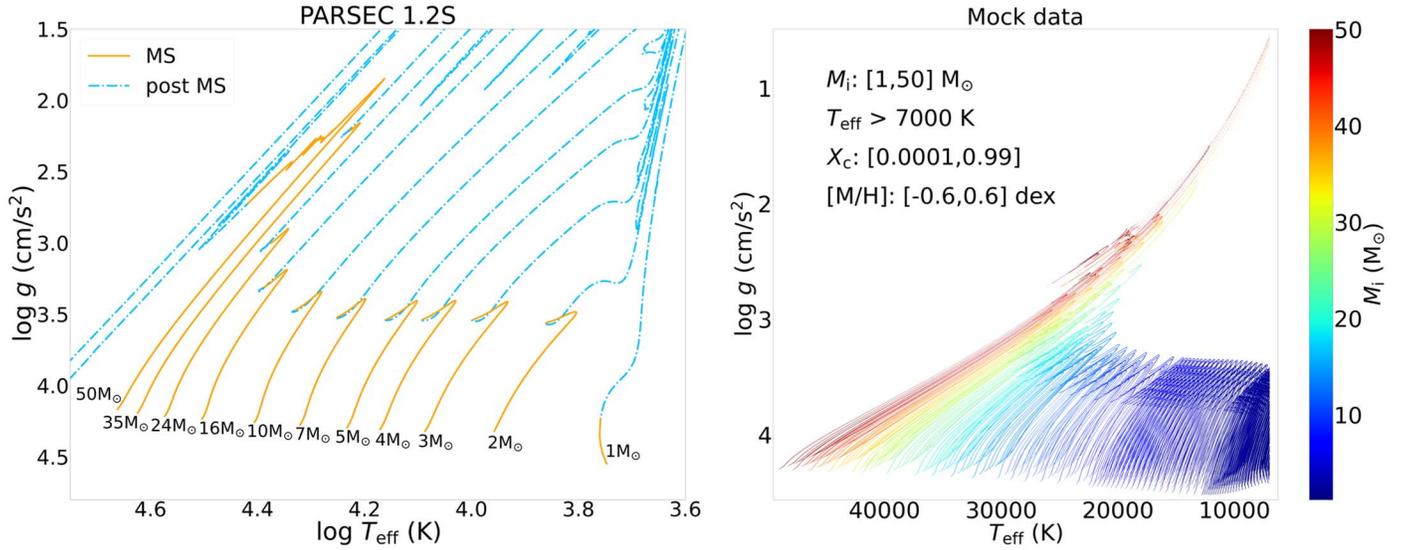

**Figure 2.** Evolutionary tracks on the log $T_{\rm eff}$–log $g$ diagram obtained from PARSEC 1.2S. The left panel shows the evolutionary tracks with different $M_i$ for $Z = 0.02$, $Y = 0.284$. The orange solid lines represent the main sequence (MS) selected based on the central hydrogen mass fraction, and the blue dashed–dotted lines shows the post–main sequence (post MS). The right panel shows the distribution of the mock data selected from PARSEC 1.2S. The different colors represent the $M_i$. The parameters of the selected sample are indicated in the top left.

**Table 2**
The Value Ranges of the Atmospheric Parameters in Our Mock Data

| $T_{\rm eff}$ (K) | log $g$ (cm s$^{-2}$) | [M/H] (dex) |
|---|---|---|
| 7000 ∼ 49,000 | 0.55 ∼ 4.52 | −0.6 ∼ 0.6 |

The tracks have $M_i$ ranging from 1 to 50 $M_\odot$ and have $T_{\rm eff}$ larger than 7000 K. The [M/H] range is set from −0.6 to 0.6 dex, slightly broader than the prediction samples range (−0.5 to 0.5 dex), to avoid inaccuracies in machine learning predictions near the boundary values. A total of 360,289 track points are chosen as the mock data, with approximately 1130 points selected for each track.

Additionally, in PARSEC 1.2S, the hydrogen mass fraction $X$ and metallicity $Z$ are given for each track. However, for the metallicity measurements, we usually obtain [M/H] from the observations. Therefore, to convert the metallicity $Z$ to [M/H], we use the following formula:

$$[{\rm M/H}] = \log((Z/X)/(Z/X)_\odot), \quad (1)$$

where the value of $(Z/X)_\odot$ is 0.0207 (A. Bressan et al. 2012). Table 2 presents the value ranges of the atmospheric parameters in our mock data. And M. Xiang et al. (2022) provided [Fe/H], which we convert to [M/H] using the following formula:

$$[{\rm M/H}] = [{\rm Fe/H}] + \log(0.638f + 0.362). \quad (2)$$

The $f = 2.91$ when [Fe/H] < −0.5; otherwise, $f = 1$ (M. Salaris et al. 1993).

### 3. Method

#### 3.1. Machine Learning Model

In this study, random forest (RF)[5] is applied to our mock data to build the model. RF is a powerful machine learning algorithm used for classification and regression tasks. It is an ensemble machine learning method that combines multiple decision trees to improve predictive accuracy. It creates decision trees based on random subsets of the data and features, which reduces overfitting and enhances generalization. This algorithm is widely used in various applications, providing robust and reliable predictions by aggregating the outputs of many individual trees.

---
[5] https://scikit-learn.org/stable/index.html

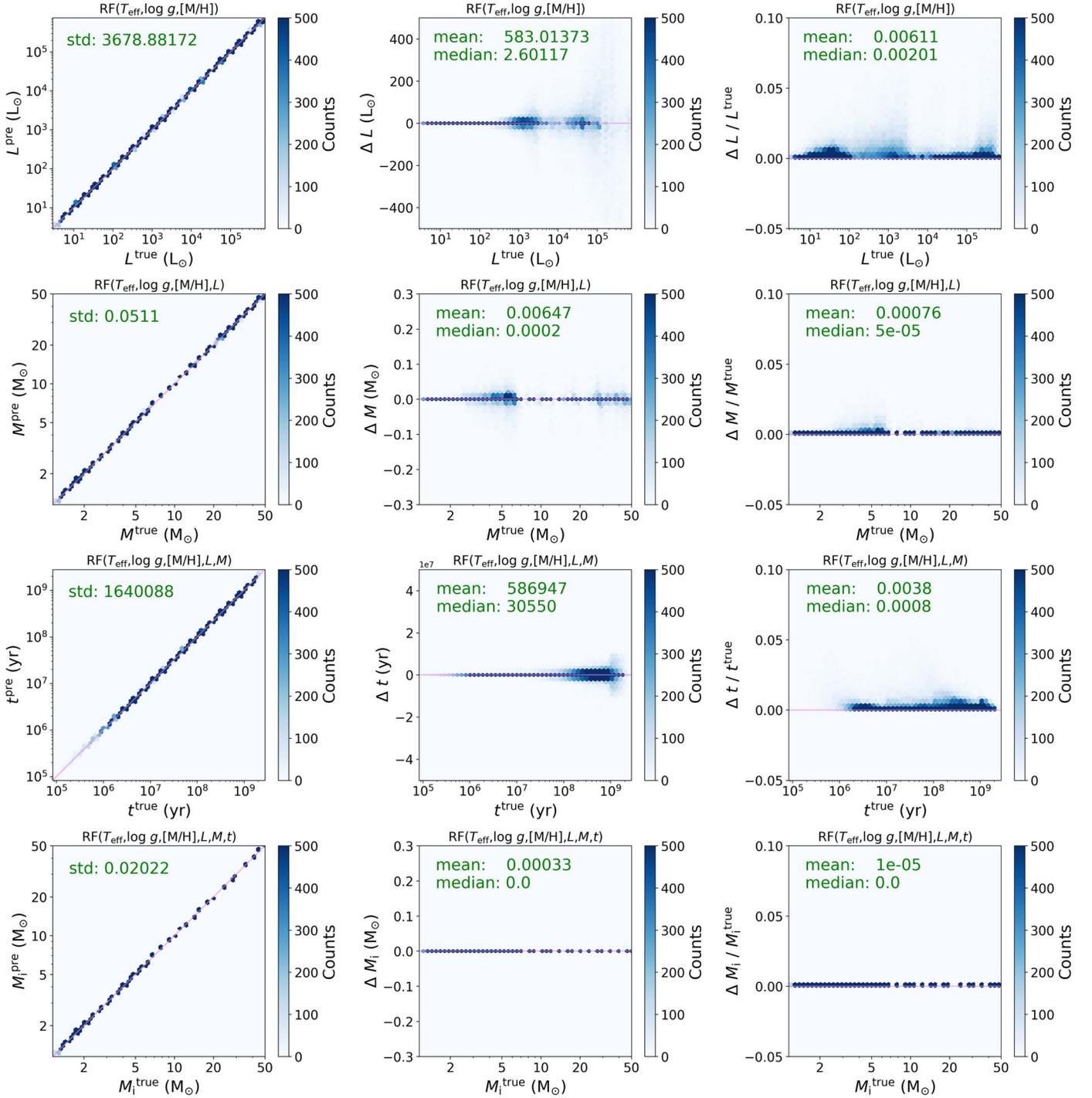

**Figure 3.** The prediction results of our trained model on the testing set. Different rows represent the predictions of $L$, $M$, $t$, and $M_i$, and from left to right, the columns indicate the direct comparisons, absolute errors, and relative errors between the true values and predictions, respectively. Each panel is labeled with the value of dispersion or the mean and median of the error in the top-left corner.

### 3.2. Model Training

During the training process, we randomly divide the mock data into a training set and a testing set with a ratio of 7:3. The testing set is used to evaluate the performance of the predictive model. The specific model training process is divided into four steps: Initially, we train the model to predict $L$ by using $T_{eff}$, $\log g$, and [M/H] as the inputs ($T_{eff}$, $\log g$, [M/H] to $L$). Subsequently, we incorporate $L$ as an additional input parameter to predict $M$ ($T_{eff}$, $\log g$, [M/H], $L$ to $M$), and $M$ is included as an input parameter when predicting $t$ ($T_{eff}$, $\log g$, [M/H], $L$, $M$ to $t$). Finally, $t$ is utilized as an input parameter for predicting $M_i$ ($T_{eff}$, $\log g$, [M/H], $L$, $M$, $t$ to $M_i$).

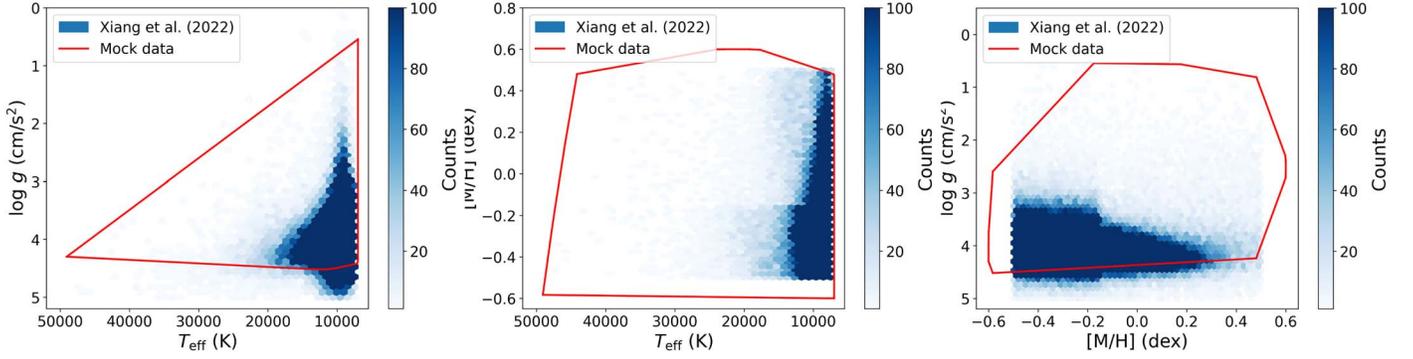

**Figure 4.** The $T_{\rm eff}$–log $g$, $T_{\rm eff}$–[M/H], and [M/H]–log $g$ distributions of the mock data and prediction sample are shown. The blue bins represent the distribution of the sample selected from M. Xiang et al. (2022), while the red wireframes indicate the ranges of $T_{\rm eff}$, log $g$, and [M/H] for our mock data from PARSEC 1.2S.

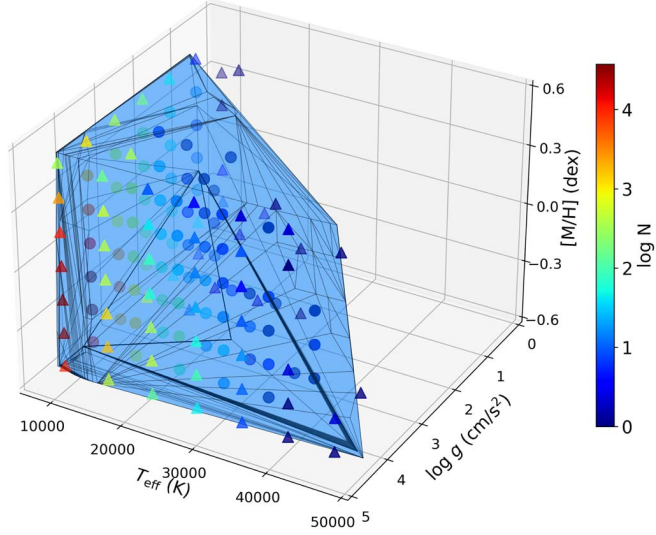

**Figure 5.** The $T_{\rm eff}$–log $g$–[M/H] distribution of M. Xiang et al. (2022) in three-dimensional space. The catalog is divided into bins in three-dimensional space, and the scatter points mean the center coordinates of each bin. The polyhedron represents the convex hull created using $T_{\rm eff}$, log $g$, and [M/H] of the mock data. There are two types of points here. The circular ones indicate they are within the convex hull, and the triangular ones indicate they are outside the convex hull. The color shows the count of each bin.

**Table 3**
The Interval Setting of $M_i$ in PARSEC 1.2S

| $M_i$ ($M_\odot$) | $\Delta M_i$ ($M_\odot$) |
|---|---|
| 1 ∼ 1.65 | 0.05 |
| 1.65 ∼ 1.95 | 0.025 |
| 1.95 ∼ 2.4 | 0.05 |
| 2.4 ∼ 6.4 | 0.2 |
| 7 ∼ 12 | 1 |
| 12 ∼ 20 | 2 |
| 20 ∼ 28 | 4 |
| 30 ∼ 50 | 5 |

To train each model, we follow these steps: first, bootstrap sampling is used to draw samples from the training set to create 2000 subsets. Next, for each subset, a decision tree is constructed, and at each node, a random subset of features is selected for splitting. This process is repeated until the 2000 decision trees are generated. Finally, the predictions from these trees are aggregated using a majority voting approach to obtain the final result. The parameter settings employed in our RF models are as follows: the maximum depth of the tree is set to unlimited; the minimum number of samples required to split an internal node is set to 2; and the minimum number of samples required to be at a leaf node is set to 1.

Figure 3 displays the validation results of the testing set, illustrating a good alignment between the predicted and true values in our model. In Figure 3, the first column shows the direct comparisons for $L$, $M$, $t$, and $M_i$. The second column demonstrates the distribution of absolute errors (pre-true) in the predictive model, and the third column displays the distribution of relative errors (|pre-true|/true). This indicates that our model can make good predictions. The discontinuity of the predicted $M_i$ is due to the fact that $M_i$ in our model is not continuous and varies in different mass gaps for different mass ranges (see Table 3).

After confirming the feasibility of our trained model, we apply it to predict $L$, $M$, $t$, and $M_i$ for early-type stars compiled from the catalogs of M. Xiang et al. (2022) and Y. Guo et al. (2021).

### 3.3. Convex Hull

The atmospheric parameters for the predicted objects may be outside the parameter ranges of the mock data, and the predicted values have relatively low reliability. Therefore, we apply the convex hull algorithm to identify stars within the parameter ranges of the mock data. Taking the catalog of M. Xiang et al. (2022) as an example, Figure 4 shows the distributions of the mock data and stars in the catalog of M. Xiang et al. (2022) on the $T_{\rm eff}$–log $g$, $T_{\rm eff}$–[M/H], and [M/H]–log $g$ diagrams. In Figure 4, the blue bins show the distributions of the stars obtained from M. Xiang et al. (2022), and the red wireframes represent the ranges of $T_{\rm eff}$, log $g$, and [M/H] for our mock data from PARSEC 1.2S. It is clear that a portion of stars falls outside the parameter ranges of the mock data.

Therefore, we create a three-dimensional convex hull using the $T_{\rm eff}$, log $g$, and [M/H] from the mock data (see Figure 5). The points represent the center coordinates of each bin for the sample from M. Xiang et al. (2022). The polyhedron is the convex hull created by the mock data. In Figure 5, we use circles to mark the stars located within the convex hull and triangles for the stars outside the convex hull. Moreover, we can observe that our mock data does not include points beyond the main sequence (see the right panel of Figure 2). However, a single convex hull cannot create a concave shape (see the left panel of Figure 4).

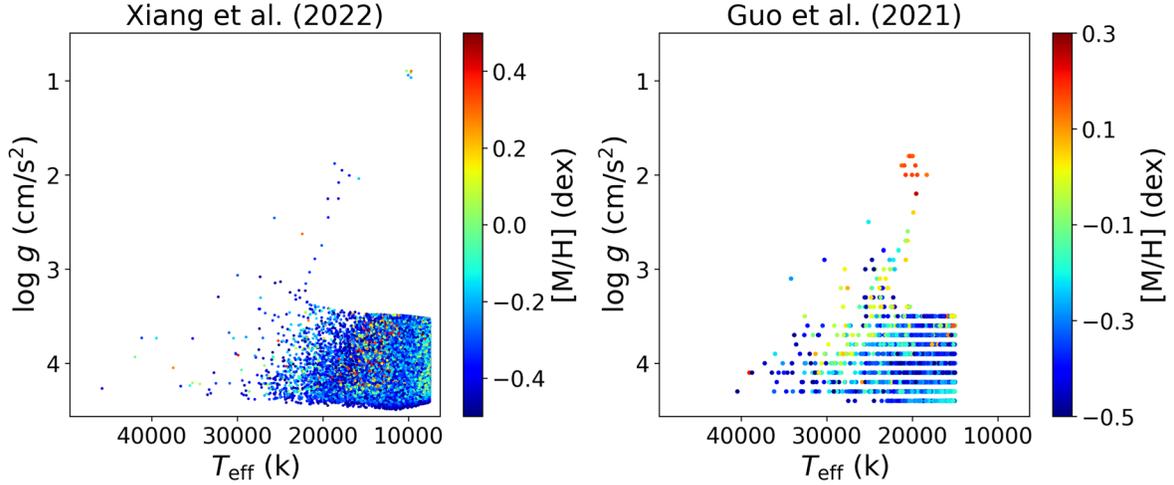

**Figure 6.** The distribution of samples filtered out by the convex hull algorithm on the $T_{\rm eff}$–log $g$ panel, with colors representing the values of [M/H].

Therefore, the first convex hull may include some stars that have evolved off the main sequence. To address this issue, we extract the points with the minimum log $g$ for each evolutionary track in the mock data, creating a second convex hull. The samples that are located in the first convex hull and not in the second convex hull are selected as the final targets.

By applying this convex hull algorithm, 132,624 stars and 3657 stars are obtained from the catalogs of M. Xiang et al. (2022) and Y. Guo et al. (2021), respectively, and are marked as "I." Figure 6 shows the distribution of samples selected by the convex hull algorithm on the $T_{\rm eff}$–log $g$ diagram for the two catalogs. It can be seen that the parameter range of these targets is consistent with the parameter range of the mock data. Then, the trained RF models are applied to derive their $L$, $M$, $t$, and $M_i$. Additionally, the uncertainties of $L$, $M$, $t$, and $M_i$ are estimated by using Monte Carlo sampling, based on the errors derived from the $T_{\rm eff}$, log $g$, and [M/H].

## 4. Results

### 4.1. Predictions of L, M, t, and $M_i$

Figure 7 demonstrates the distribution of the predicted $L$, $M$, $t$, and $M_i$ in Galactic coordinates, and it shows that stars in the low-Galactic-latitude region have larger values of $L$ and $M$, but smaller values of $t$. The region near the Galactic disk is the major star-forming region, with a large number of molecular clouds that favor the stellar formation. Consequently, these stars are relatively young, massive, and have not yet evolved. And there is a clear mass–luminosity relation ($M$–$L$ relation) in early-type stars in the main-sequence stage.

We examine the $M$–$L$ relation of our prediction and compare it with the $M$–$L$ relations in previous studies (Table 4). Figure 8 shows the $M$–$L$ relations obtained from our study (orange solid line). For comparison, the results from S. C. Griffiths et al. (1988; green dashed line), O. Demircan & G. Kahraman (1991; red dotted line) and Z. Eker et al. (2015; purple dashed–dotted line) are also presented. We fit the $M$–$L$ relation using the least-squares method. The $M$–$L$ relation derived by our sample is

$$L/L_\odot \approx 2.544(M/M_\odot)^{3.665}. \quad (3)$$

In addition, we plotted the distribution of the main-sequence-age-to-lifetime ratio for stars of different masses, as shown in Figure 9. The result shows that the number of old stars is generally higher than that of young stars. Figure 10 displays the distribution of our sample and mock data on the $T_{\rm eff}$–log $L$ panel. In the left panel of Figure 10, the two black lines show the boundary range of the mock data. It can be seen that our predicted parameters fall well within the range of the mock data parameters. And the other different color lines represent the various evolutionary stages with solar metallicity within the main sequence.

### 4.2. The Differences between the Two Prediction Samples

Furthermore, we compare the differences in our predicted $L$, $M$, $t$, and $M_i$ between the two prediction samples. First, we conduct a crossmatch between the two prediction samples and obtain 1206 common stars that are flagged as "I" by us, within the credible range of their catalogs (refer to Section 2.1), and with good spectral S/N (S/N > 30). Subsequently, we compare their $L$, $M$, $t$, and $M_i$ in Figure 11. As Figure 11 shows, the median relative differences (|Xiang − Guo|/Guo) of $L$, $M$, and $M_i$ are 29%, 7%, and 7%. Compared to $L$, $M$, and $M_i$, $t$ exhibits a relatively larger difference between the two samples, with the value of 36%. And the systematic bias observed in each comparison is caused by $T_{\rm eff}$.

We then compare the differences in $T_{\rm eff}$, log $g$, and [M/H] between the two samples in Figure 12. As Figure 12 shows, for the two catalogs, the standard deviations of $T_{\rm eff}$, log $g$, and [M/H] are 1420 K, 0.20 dex, and 0.20 dex, respectively. These differences may contribute to the discrepancies in the estimates of their $L$, $M$, $t$, and $M_i$.

## 5. Discussion

### 5.1. External Validation

W. Sun et al. (2021) has provided a catalog of 40,034 late-B and A-type stars, and the atmospheric parameters of these stars are obtained from the LAMOST medium-resolution spectra using the SLAM method. The credible range for this catalog is 6000 K ⩽ $T_{\rm eff}$ ⩽ 15,000 K, 3.5 dex ⩽ log $g$ ⩽ 4.5 dex, and −1 dex ⩽ [M/H] ⩽ 1 dex. And the uncertainty of $T_{\rm eff}$, log $g$, and [M/H] is 75 K, 0.06 dex, and 0.05 dex, respectively, for the stars with an S/N greater than 60. Additionally, the log $L$ is derived by combining the photometric data from Gaia and the Two Micron All Sky Survey, while the $M$ is derived by fitting $\log T_{\rm eff}$, $\log L$, and [M/H] using the PARSEC 1.2S evolutionary tracks. And the uncertainty of $\log L$ and $M$ is 0.02 dex and 0.1 $M_\odot$, respectively. Then, we compare our $L$ and $M$ with the results presented by W. Sun et al. (2021).

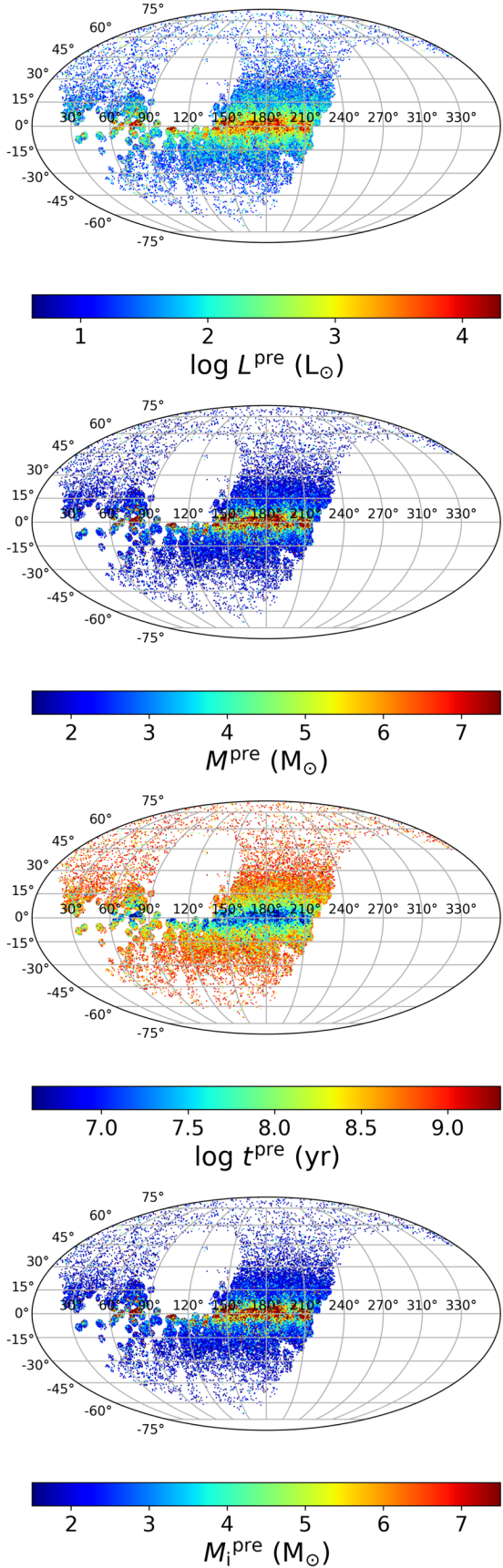

**Figure 7.** The distribution of predicted $L$, $M$, $t$, and $M_i$ for our sample in the Milky Way. Different panels represent different predicted parameters, and the color indicates the value of each parameter.

**Table 4**
List of $M$–$L$ Relations from Different Studies

| Reference | Mass Range ($M_\odot$) | $M$–$L$ Relation |
|---|---|---|
| S. C. Griffiths et al. (1988) | $M < 0.40$ | $L/L_\odot \approx 0.30(M/M_\odot)^{2.44}$ |
|  | $0.40 < M < 5.01$ | $L/L_\odot \approx 1.01(M/M_\odot)^{4.16}$ |
|  | $M > 5.01$ | $L/L_\odot \approx 2.34(M/M_\odot)^{3.51}$ |
| O. Demircan & G. Kahraman (1991) | $0.10 < M < 18.10$ | $L/L_\odot \approx 1.18(M/M_\odot)^{3.70}$ |
|  | $M < 0.70$ | $L/L_\odot \approx -0.35(M/M_\odot)^{2.62}$ |
|  | $M > 0.70$ | $L/L_\odot \approx 1.02(M/M_\odot)^{3.92}$ |
| Z. Eker et al. (2015) | $0.38 < M \leqslant 1.05$ | $L/L_\odot \approx 0.94(M/M_\odot)^{4.84}$ |
|  | $1.05 < M \leqslant 2.40$ | $L/L_\odot \approx 1.00(M/M_\odot)^{4.33}$ |
|  | $2.40 < M \leqslant 7.00$ | $L/L_\odot \approx 1.32(M/M_\odot)^{3.96}$ |
|  | $7.00 < M \leqslant 32.00$ | $L/L_\odot \approx 17.26(M/M_\odot)^{2.73}$ |

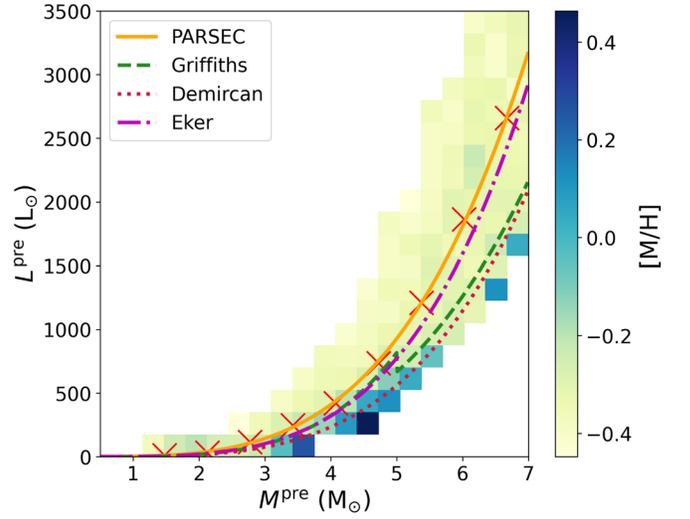

**Figure 8.** The $M$–$L$ relations of the stars from our sample. The red crosses represent the median values of the predicted $L$ in each mass bin, and the orange curve represents the $M$–$L$ relation from PARSEC 1.2S, fitted by the least-squares method. The curves in other styles represent the work of others from Table 4. The color indicates the value of [M/H].

Initially, we select the stars within the credible range of W. Sun et al. (2021) and then remove the stars from the catalog of M. Xiang et al. (2022) that are not flagged as "I" by us. Subsequently, we conduct a crossmatch between the two catalogs, resulting in 5176 common stars with an S/N greater than 30 for comparison. The comparison results are illustrated in Figure 13. As Figure 13 shows, the $x$-axis represents the values from W. Sun et al. (2021), and the $y$-axis represents our predicted values for the sample selected from M. Xiang et al. (2022). The left panel displays the comparison of $L$, while the right panel presents the comparison of $M$. The results indicate good consistency. They illustrate that the median relative errors (|pre-Sun|/Sun) of $L$ and $M$ are 32% and 9%, respectively. And it is worth noting that this level of difference is comparable to the typical error observed when directly predicting values using stellar evolutionary tracks (A. Serenelli et al. 2021). However, it can be seen that there is a systematically lower value for *M* compared to W. Sun et al. (2021). This discrepancy may be attributed to differences in the [M/H] between the two catalogs (as shown in Figure 14).

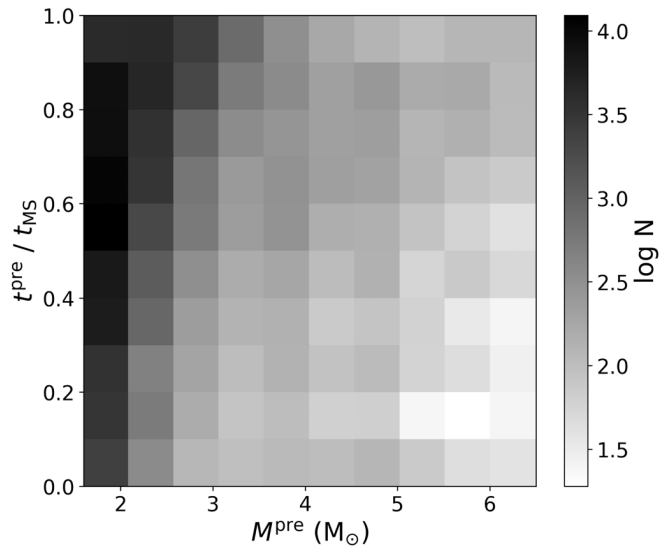

**Figure 9.** The distribution of the ratio of the lifetime for stars of different masses for our sample. The *x*-axis represents the predicted *M*, and the *y*-axis represents the predicted *t* divided by the main-sequence lifetime. The color indicates the count in log of each bin.

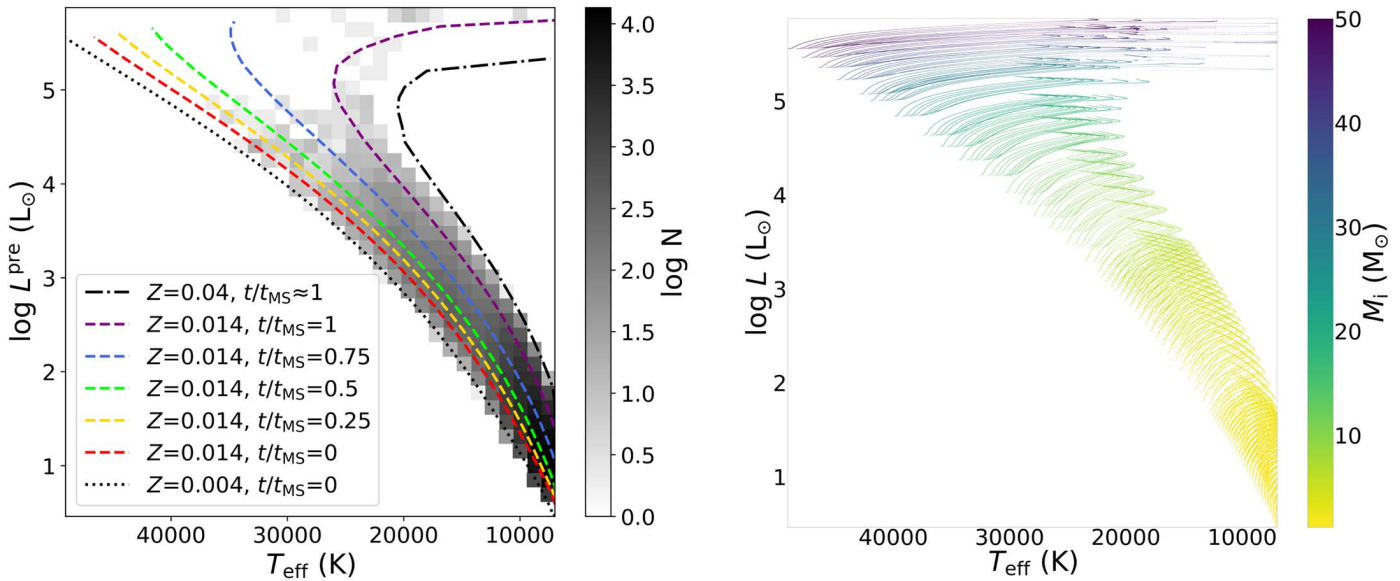

**Figure 10.** The distribution on $T_{\rm eff}$–log *L* for our sample (left) and mock data (right). In the left panel, the two black lines show the boundary range of the mock data. The other colored lines represent the different degrees of evolution of stars with solar metallicity in the main sequence. In the right panel, the color represents the $M_i$.

For further validation, we apply our model to the catalogs of W. Sun et al. (2021) and G. Kordopatis et al. (2023) on their atmospheric parameters. We compare our predicted *M*, *t*, and $M_i$ with their results. W. Sun et al. (2021) utilized PARSEC to predict the *M* of early-type stars, while G. Kordopatis et al. (2023) provided a catalog of ∼5 million stars with *t* and $M_i$ labels by using PARSEC to fit atmospheric parameters. For the catalog of G. Kordopatis et al. (2023), stars with $M_i$ greater than 1 $M_\odot$, $T_{\rm eff}$ higher than 7000 K, and [M/H] within the range of −0.5 to 0.5 dex are selected. The comparison results are shown in Figure 15. They show good consistency in *M* and $M_i$ predictions, but the difference in *t* prediction is slightly larger. And it is noteworthy that the uncertainty of *t* and $M_i$ is 2 Gyr for giants and main-sequence stars and 0.1 $M_\odot$ for main-sequence and turn-off stars in G. Kordopatis et al. (2023).

In addition, Bayesian isochrone fitting (B. R. Jørgensen & L. Lindegren 2005) is a widely used methodology for estimating the *M* and *t* of stars. The main advantage of an RF-based approach is its speed and efficiency once the model is trained. It can predict multiple points simultaneously while achieving a prediction accuracy comparable to that of Bayesian isochrone fitting. For comparison, 3000 random samples are selected from the testing set. Then, we apply the trained RF model and Bayesian isochrone fitting to these samples. The results of the comparison are presented in Figure 16. The *x*-axis represents the true values of the 3000 samples, while the *y*-axis displays the predicted results from Bayesian isochrone fitting (first row) and RF (second row). For the Bayesian isochrone fitting, the relative differences (|pretrue|/true) between the predicted values of $L$, $M$, $t$, and the true values are 0.02%, 0.01%, and 0.21%, respectively, whereas for RF, the relative differences are 0.21%, 0.006%, and 0.08%, respectively. These results indicate that RF achieves comparable accuracy to that of Bayesian isochrone fitting. Notably, RF can complete predictions for these 3000 samples on an eight-core machine in under 3 hr, while Bayesian isochrone fitting requires approximately 3 days. Additionally, F. Anders et al. (2023) also compared Bayesian isochrone fitting with machine learning. While Bayesian isochrone fitting can improve computational speed by reducing the density of the isochrones, it still requires significant time. In contrast, machine learning methods produce predictions comparable to Bayesian isochrone fitting while requiring much less time. Therefore, it is evident that the RF method not only ensures the same accuracy as Bayesian isochrone fitting but also saves a considerable amount of time and is more suitable for the measurement of stellar parameters in large samples.

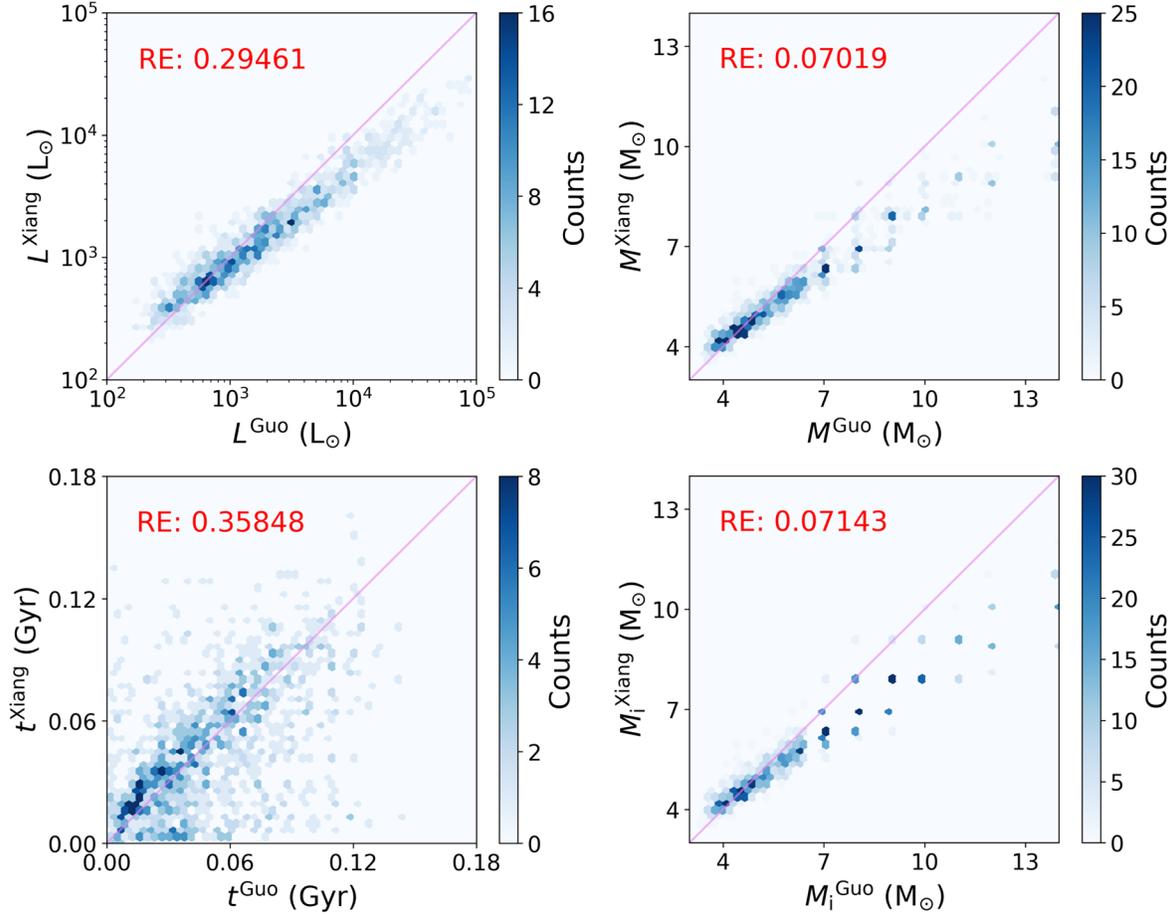

**Figure 11.** The differences in our predicted values of $L$, $M$, $t$, and $M_i$ between the two prediction samples. The *x*-axis represents the predicted values for Y. Guo et al. (2021), and the *y*-axis is those from M. Xiang et al. (2022). The median relative error has been marked at the top left of each panel.

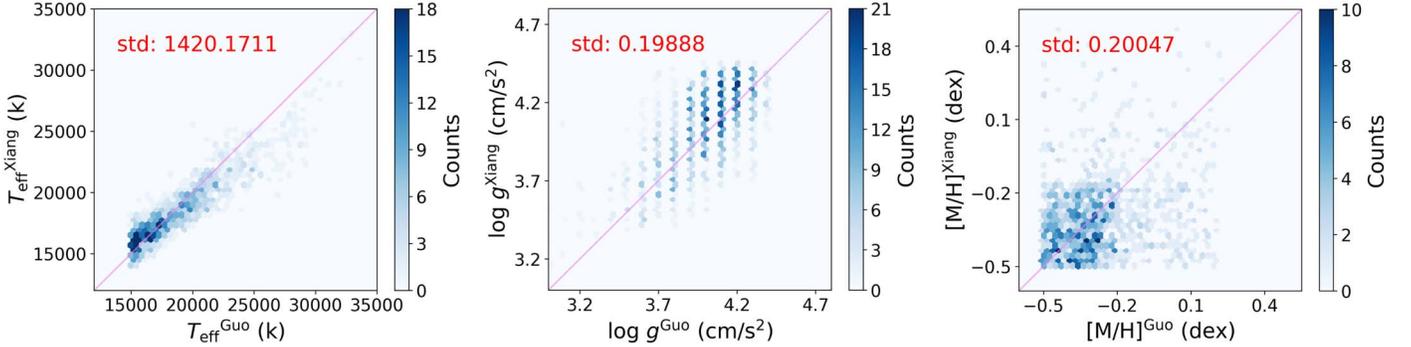

**Figure 12.** The atmospheric parameters between the two samples we used. The *x*-axis represents the atmospheric parameters from Y. Guo et al. (2021), and the *y*-axis is those from M. Xiang et al. (2022).

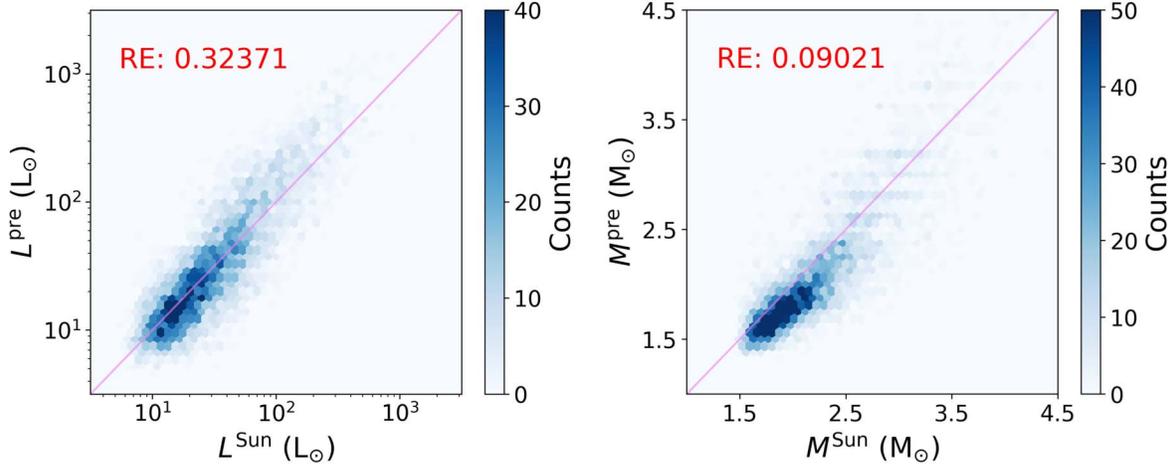

**Figure 13.** Comparison of $L$ and $M$ with that of W. Sun et al. (2021). The x-axis represents the values from W. Sun et al. (2021), and the y-axis represents our predicted values for the sample selected from M. Xiang et al. (2022). The median relative error is indicated at the top left of each panel.

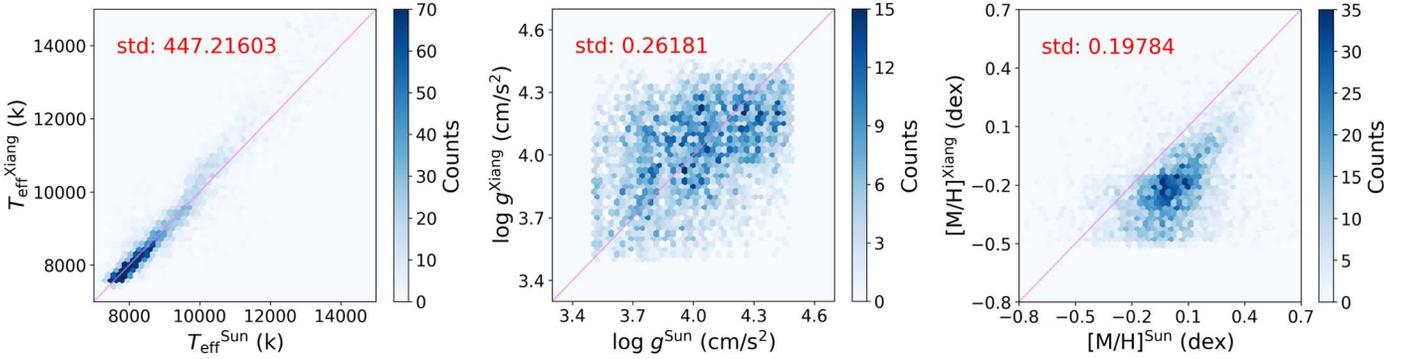

**Figure 14.** The difference in atmospheric parameters between M. Xiang et al. (2022) and W. Sun et al. (2021). The x-axis represents the values from W. Sun et al. (2021), and the y-axis represents the values from M. Xiang et al. (2022).

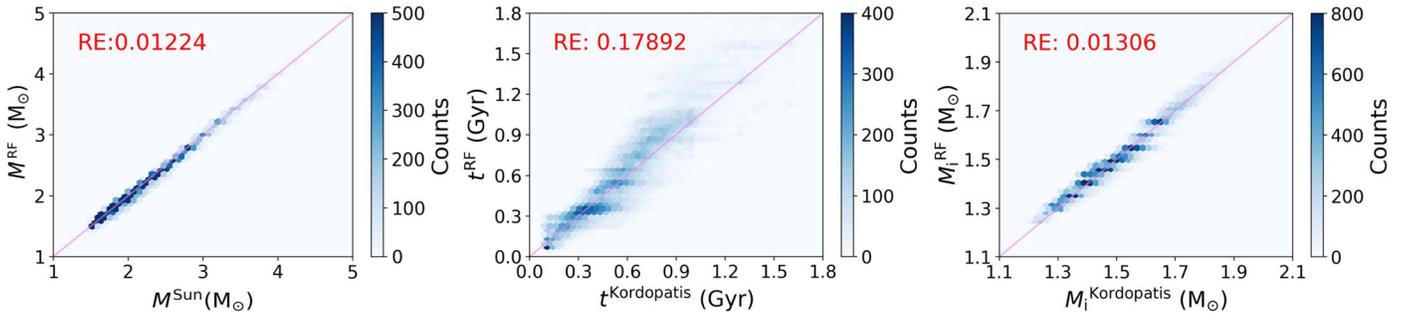

**Figure 15.** Comparison with the literature. The x-axis represents the literature values, and the y-axis represents the predictions of our model. The left panel shows the $M$ comparison with W. Sun et al. (2021), while the middle and right panels show the $t$ and $M_i$ comparisons, respectively, with G. Kordopatis et al. (2023). The median relative error is marked on the top left of each panel.

### 5.2. Validating L with Gaia

Additionally, the comparison between our predicted $L$ and Gaia's observations is shown in Figure 17. In Figure 17, Gaia's observations are converted to $L$ by Equation (4):

$$m_G = M_G + DM$$
$$m_G = M_{bol} - BC_G + DM$$
$$m_G = 4.75 - 2.5 \log L - BC_G + DM$$
$$\log L = (4.75 - m_G - BC_G + DM) / 2.5. \quad (4)$$

Here, DM is the distance modulus ($DM = 5 \log D - 5$) derived from C. A. L. Bailer-Jones et al. (2021). BC is the bolometric correction adopted from the Python package "isochrones,"[6] and the extinction correction can be applied to $BC_G$ by simply inputting the $A_V$ values ($A_V = 3.16 \times E(B - V)$, where 3.16 was from S. Wang & X. Chen (2019) and $E(B - V)$ was obtained from G. M. Green et al. 2019). The comparison is restricted to stars with good spectral ($S/N > 30$) and good parallax measurements ($e_{plx}/plx < 0.2$). Figure 17 indicates a good alignment with a relative difference of 36%.

---

[6] https://isochrones.readthedocs.io/en/latest/index.html

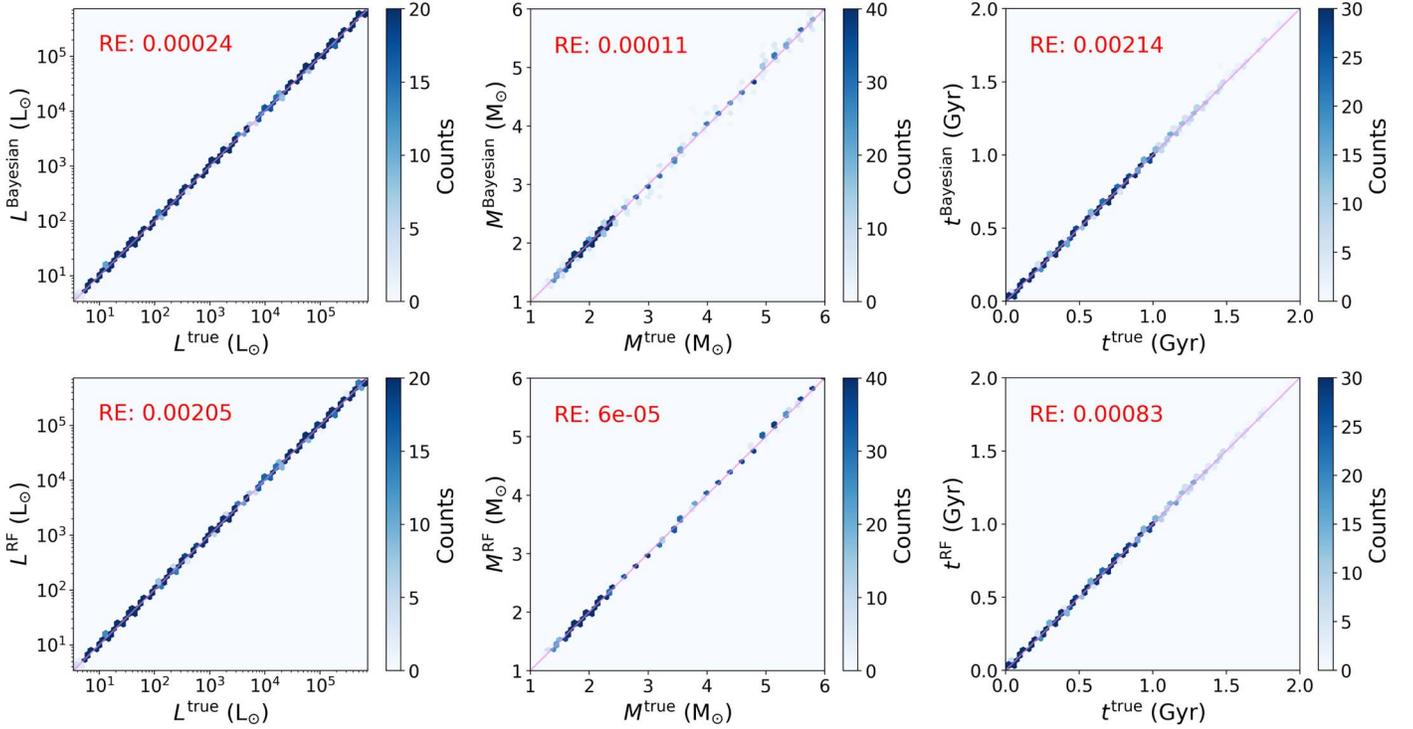

**Figure 16.** Comparison of prediction results between Bayesian isochrone fitting and RF on the testing set. The *x*-axis represents the true values from the testing set, while the *y*-axis shows the predicted results from Bayesian isochrone fitting (first row) and the RF model (second row).

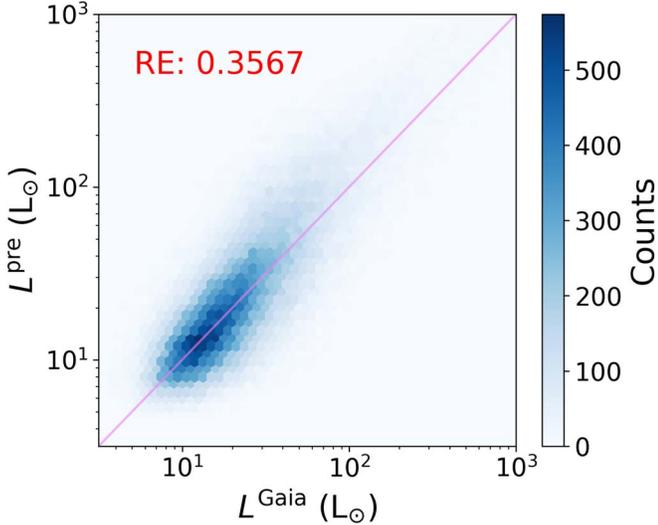

**Figure 17.** Comparison between our predicted *L* and Gaia's observations. The values on the *y*-axis represent our predicted *L*, while the values on the *x*-axis are obtained through the conversion of Gaia DR3.

### 5.3. Validating Stellar Ages with Wide Binaries

We use wide binaries to validate our predicted $t$. Binary stars are usually considered to be born at the same time, therefore, they also have the same age and the same chemical components. First, we obtain 92 matched wide binaries from H.-C. Hwang et al. (2022). These wide binaries are flagged as "I" by us and have S/N > 30. The comparison result is shown in Figure 18. The measurement error is 0.1 dex for [Fe/H] measured by M. Xiang et al. (2022), we categorize the points into two types based on the 0.1 dex. The x-shaped markers are the samples with $\Delta$[Fe/H] > 0.1, and the triangles represent the samples with $\Delta$[Fe/H] ⩽ 0.1. It can be seen that there is an obvious dispersion with the value of 0.44 Gyr for the 92 wide binaries, and a better alignment of the samples with $\Delta$[Fe/H] ⩽ 0.1. We then compare the differences in [Fe/H] among these 92 wide binaries, and found that the consistency of [Fe/H] in wide binaries is not very good to begin with. It indicates that the inaccuracy in the *t* prediction may be due to the measurement error of metallicity, as metallicity has a significant impact when predicting *t*. Therefore, in Section 5.4 we discuss how metallicity affects *t* prediction.

### 5.4. The Impact of Metallicity on the Model

We investigate the impact of metallicity on the prediction of the four parameters *L*, *M*, *t*, and $M_i$. We conduct two sets of tests: one using incorrect [M/H] values to make predictions on the testing set, and the other without using [M/H]. The results from both sets are then compared with the predictions made using the correct [M/H].

In Figure 19, the *x*-axis represents the predictions made using the correct [M/H]. The first row shows the results using incorrect [M/H] (with 0 as the input). The second row represents the case where [M/H] is not used. In this figure, the $\Delta$ values represent the differences between predictions with [M/H] = 0 (or without [M/H]) and those with correct [M/H] (|Prediction$^{[M/H]:0}$ − Prediction$^{[M/H]:correct}$| or |Prediction$^{[M/H]:without}$ − Prediction$^{[M/H]:correct}$|, where the "Prediction" means the *L*, *M*, *t*, and $M_i$). As we can see, the first row shows that the incorrect [M/H] leads to an increased measurement error for predictions. Then, the second row shows that when [M/H] is not used in the model training, the relative errors of *L*, *M*, and $M_i$ are obviously reduced compared to the first row, but not for *t*.

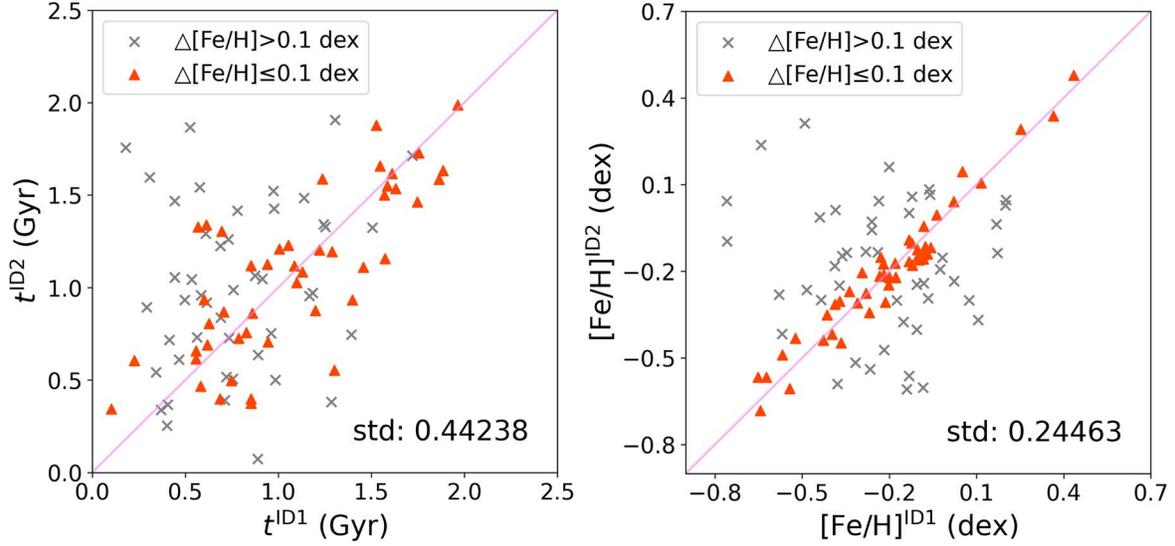

**Figure 18.** Comparison of $t$ (left) and [Fe/H] (right) between the two components in wide binaries. The $t$ is predicted by our method, and the sample of wide binaries is obtained from H.-C. Hwang et al. (2022). The points are categorized into two types based on the $\Delta$[Fe/H]. The values of [Fe/H] are from the catalog of M. Xiang et al. (2022).

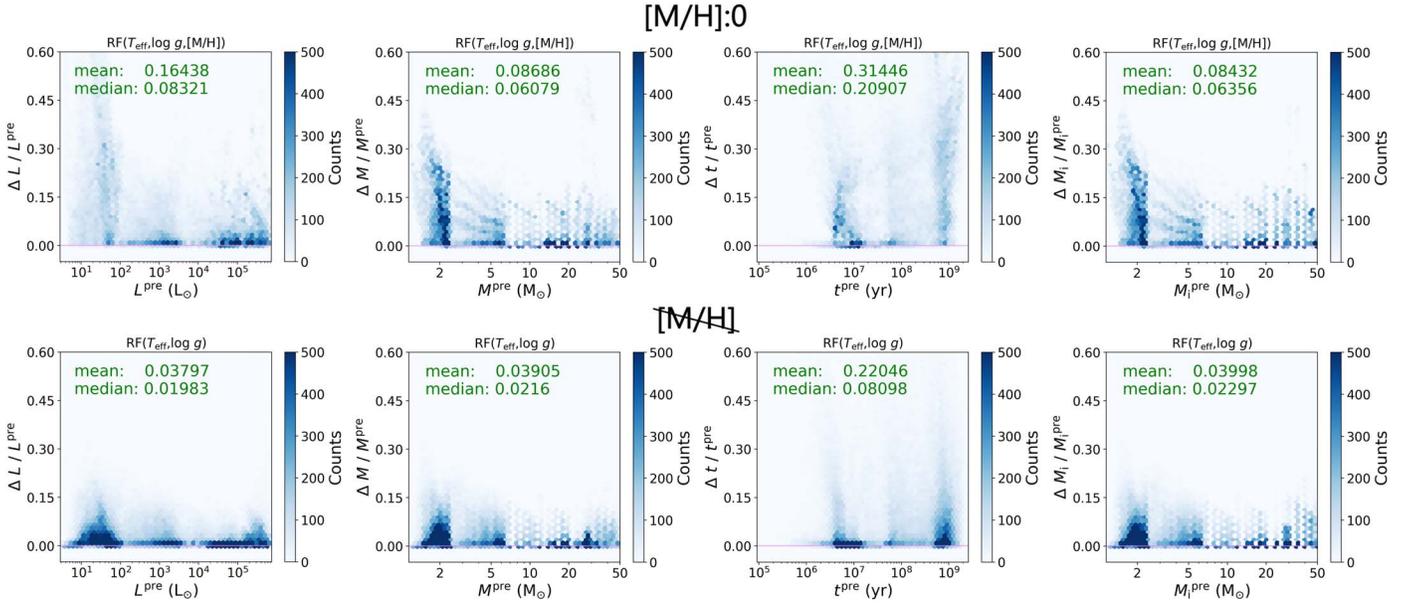

**Figure 19.** The impact of metallicity on predictions in the testing set. Different columns represent the relative error distribution of the $L$, $M$, $t$, and $M_i$. The first row shows the prediction with an incorrect [M/H] value input, and the second row shows the prediction without using [M/H]. The x-axis represents the predictions made using the correct metallicity values.

We then compare the differences between $L$, $M$, $t$, and $M_i$ with the inputs of $T_{\rm eff}$, $\log g$, and [M/H] (x-axis) and the inputs of $T_{\rm eff}$ and $\log g$ (y-axis). Figure 20 shows the comparison results for M. Xiang et al. (2022) and Y. Guo et al. (2021). It can be seen that the impact of metallicity on $t$ is the greatest, with a difference of 25%. For $L$, $M$, and $M_i$, the differences are all around 6%.

## 6. Conclusion

In this paper, we train a machine learning model by using the evolutionary tracks in the main-sequence stage from PARSEC 1.2S to obtain the $L$, $M$, $t$, and $M_i$ for main-sequence early-type stars. Based on the derived atmospheric parameters of early-type stars from M. Xiang et al. (2022) and Y. Guo et al. (2021), we predict their $L$, $M$, $t$, and $M_i$. Since the parameter distribution of predicted catalogs exceeds the parameter distribution of the mock data, to better use the catalogs we provided, we use the convex hull algorithm to select stars with high-confidence predictions and flag them with "I."

Additionally, we then compare our predicted $L$ and $M$ with these derived from W. Sun et al. (2021) and found good consistency with differences of 32% and 9%, respectively. When comparing $L$ with Gaia $G$-band photometric observation, a discrepancy of 36% is observed. Next, we validate our predicted $t$ using the wide binaries catalog of H.-C. Hwang et al. (2022) and found a dispersion of 0.44 Gyr. The dispersion may stem from errors in metallicity measurements. Finally, we investigate the impact of metallicity on the prediction of $L$, $M$, $t$, and $M_i$. The results show that its impact on $t$ is the greatest.

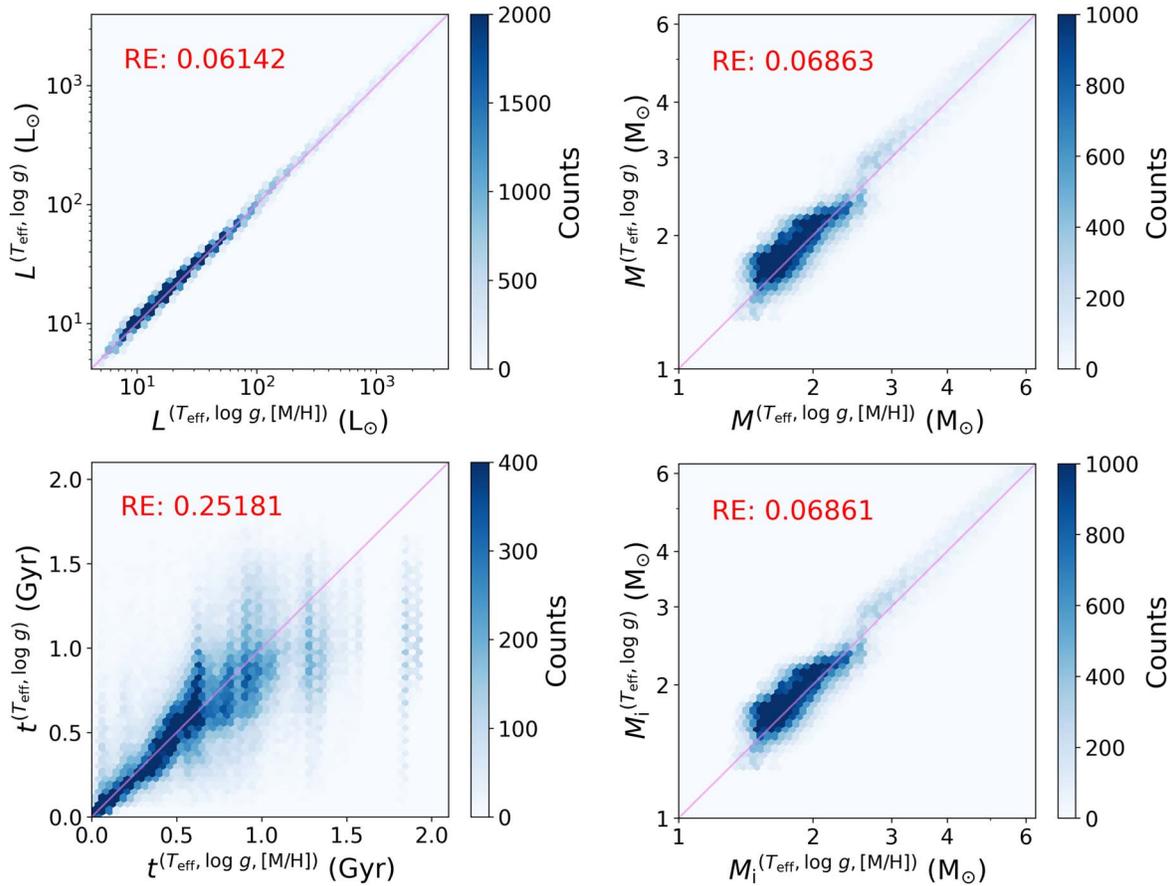

**Figure 20.** The impact of metallicity on predictions in our sample. Different panels show the predictions of $L$, $M$, $t$, and $M_i$. The $x$-axis represents the prediction with [M/H], while the $y$-axis represents the prediction without [M/H].

The two predicted catalogs are available for download from the online catalogs.tar.gz archive. The study of early-type stars in the Milky Way has significant scientific importance, and our work will be extended in the future.

### Acknowledgments

This work is supported by the National Natural Science Foundation of China (grant Nos. 12288102, 12125303, 12090040/3), the National Key R&D Program of China (grant Nos. 2021YFA1600401, 2021YFA1600403, 2021YFA1600400), the Natural Science Foundation of Yunnan Province (Nos. 202201BC070003, 202001AW070007), the International Centre of Supernovae, Yunnan Key Laboratory (No. 202302AN360001), the Yunnan Revitalization Talent Support Program—Science & Technology Champion Project (No. 202305AB350003), the China Manned Space Project (No. CMS-CSST-2021-A10), the NFSC (grant No. 12303106), and the Postdoctoral Fellowship Program of CPSF (No. GZC20232976).

### ORCID iDs

Jianping Xiong ◎ https://orcid.org/0000-0003-4829-6245
Jiao Li ◎ https://orcid.org/0000-0002-2577-1990
Yanjun Guo ◎ https://orcid.org/0000-0001-9989-9834
Zhanwen Han ◎ https://orcid.org/0000-0001-9204-7778
Xuefei Chen ◎ https://orcid.org/0000-0001-5284-8001
Chao Liu ◎ https://orcid.org/0000-0002-1802-6917